\documentstyle[12pt]{article}
\include{amssymb}
\begin{document}
\newcommand{\be}{\begin{equation}}
\newcommand{\ee}{\end{equation}}
\title{On the triviality of certain non-Riemannian models of gravitation}
\author{R. Scipioni}
\maketitle
Department of Physics and Astronomy, The University of British Columbia,\\
6224 Agricultural Road, Vancouver, B.C., Canada V6T 1Z1 \footnote{scipioni@physics.ubc.ca}
\bigskip
\bigskip
\bigskip
\bigskip
\bigskip
\begin{abstract}
We prove in the Tucker-Wang approach to non-Riemannian Gravity that a general homogeneous Lagrangian density in the general connection with order of homogeneity of at least two gives no contribution to the generalised Einstein equations. Other important cases are also considered.\\
\\
04.20.-q, 04.40.-b, 04.50.+h, 04.62.+v
\end{abstract}
\newpage
\section{Introduction and Motivation}
Recently much effort has been devoted to the study of some non-standard gravitational theories, that is theories which allow for non-metricity and torsion of the spacetime. The usual approach to this generalization of the Einstein theory goes through the gauge field method which permits to obtain a gauge theory of gravity starting from the affine group $A(n, R)$ [1].\\
In this approach the metric $g_{ab}$ the connection $\omega^{a}{}_{b}$ and the coframe $e^{a}$ are considered as three independent gauge potentials whose fields are the non-metricity $Q_{ab}$ the curvature $R^{a}{}_{b}$ and the torsion respectively $T^{a}$.\\
However when a detailed study is performed we note that the different equations we get are not independent, in particular the one for the coframe and the one for the metric are related, meaning that the approach contains a kind of redundancy [1].\\
It has been suggested by Tucker-Wang [2,3] to drop one of the potentials like the metric or the coframe and use only the connection and the metric, or the coframe as independent variables in a pure variational approach. This approach is motivated also by the fact that when we describe the symmetry reduction from the general affine group, to the group describing the low energy limit of gravity we still have the freedom of choosing the coframe. This permit us to choose an orthonormal coframe and by doing so, the degree of freedom of the metric and the coframe becomes equivalent. The metric can be written as: $g = \eta_{ab} e^{a} \otimes e^{b}$ with $\eta_{ab} = diag(-1,1,1,1,...)$.\\
In this approach we choose the coframe $e^{a}$ and the connection as independent variables to obtain the field equations using a variational technique.\\
Recently this approach has been used to prove that a remarkable reduction property occurs in the field equations of certain non-Riemannian models of gravity [4]. This property of reduction has been proved in Metric-Affine gravity too [5,6] and it is now known as the Dereli-Obukhov-Tucker-Wang theorem [5,9].\\
This theorem has been proved mainly with computer programs able to perform intense calculations in differential geometry.\\
In this paper we give a formal proof of a number of results which are consequence of the homogeneity properties of the action density and the Cartan equation.\\
Invoking the relation between the Cartan equation and the generalised Einstein equations we are able to prove that some contributions in the generalised Einstein equations may vanish identically when the solution for the non-Riemannian part of the connection is substituted into them.  

\section{Formulation of the problem and main result}
\bigskip
Consider a generic action dependent on the variables $e^{a}$ and $\lambda^{a}{}_{b}$:
\be
S = \int L(\lambda, e)
\ee
where $e^{a}$ and $\lambda^{a}{}_{b}$ indicates the coframe and the non-Riemannian part of the connection.\\
We will suppose $L(\lambda,e)$ to be homogeneous of order $n \ge 2$ in $\lambda$.\\
The mathematical formulation of the condition of homogeneity is:
\be
\lambda^{a}{}_{b} \wedge \frac{\delta}{\delta {\lambda^{a}{}_{b}}} L(\lambda,e) = n L(\lambda,e)
\ee
The field equations are obtained by considering independent variations with respect to $\lambda^{a}{}_{b}$ and the coframe $e^{c}$.\\
The equation obtained from the $\lambda$ variation is defined to be the ${\bf Cartan}$ Equation:
\be
\frac{\delta}{\delta \, \lambda} L(\lambda,e) \equiv Cartan(\lambda ,e) \equiv 0
\ee
While the equation obtained from the Coframe variation is the Generalised Einstein equation:
\be
\frac{\delta}{\delta e^{c}} L(\lambda,e) \equiv Einstein(\lambda,e) = 0
\ee
From the Cartan equation is possible to solve for $\lambda^{a}{}_{b}$ so that:
\be
Cartan({\tilde{\lambda}}_{ab},e) = 0
\ee
The question arises as to whether the Einstein equation have special properties when the solution for $\lambda^{a}{}_{b}$ is used in the generalised Einstein equations:
\be
Einstein({\tilde{\lambda}}_{ab},e) \equiv 0
\ee
\\
Then we have:\\
\\
{\bf Definition 1:} \emph{The theory is said to be trivial if the previous relation is satisfied. That meaning that it is not an equation to be solved for the metric (coframe) but an identity}\\
\\
Our main result then can be stated as follows:\\
\\
{\bf Theorem 1:} \\
\\
\emph{Suppose that} $L(\lambda,e)$ \emph{is homogeneous of at least order two in} $\lambda$ \emph{and contains only two independent variables} $\lambda,e$. \emph{Then the theory is trivial in the sense that} $Einstein(\tilde{\lambda},e)$ \emph{vanish identically}\\
\\ 
From this theorem which will be proven in the next section it follows that any homogeneous Lagrangian density in $\lambda$ gives a trivial Einstein equations.\\
As an example of application consider the following Lagrangian density:
\be
L(\lambda,e) = k R \star 1 + \frac{\beta}{2} (Q \wedge \star Q)  + \frac{\gamma}{2} (T \wedge \star T)
\ee
Where $k,\beta, \gamma$ are constants, and:
\be
Q = - 2 \lambda^{a}{}_{a} \, \, \, T= i_{a} T^{a} \, \, \, T^{a} = \lambda^{a}{}_{b} \wedge e^{b}
\ee
where $i_{a}$ is the contraction operator defined as $i_{a}(e^{b}) = \delta^{b}{}_{a}$.
We can write:
\be
R \star 1 = \stackrel{0}{R} \star 1 + \Delta \stackrel{0}{R} \star 1 
\ee
Where $\stackrel{o}{R} \star 1$ is the Levi-Civita Einstein Hilbert term.\\
It can be seen that apart from a total derivative $\Delta \stackrel{0}{R} \star 1$ is bilinear in $\lambda$, then by using the theorem, we can certainly say that the Einstein equations obtained from the previous Lagrangian density reduce to the Einstein equations coming from the term $\stackrel{o}{R} \star 1$, that is the Riemannian Einstein equations of General Relativity. This can be verified by direct calculations.\\
In the next section we will see how using a generalised version of the theorem the result for the action (7) can be generalised to more general actions.\\
\newpage
\section{The Proof}
\bigskip
The first elementary property which we have to use is the following:\\
\\
{\bf Lemma 1 :} \emph{The variations with respect to the coframe and the connection are independent so that} $\delta_{\lambda} e = 0$ ; $\delta_{e} \lambda = 0$ \emph{and} $\delta_{e} \delta_{\lambda} A(\lambda ,e) = \delta_{\lambda} \delta_{e} A (\lambda,e)$.\\
\\
{\bf Proof} \\
It is an immediate consequence of the fact that any multi-linear object dependent on $\lambda$ and $e$ can be expanded as functions which contain $\lambda$ and $e$ as distinct objects. Then from the assumed independence we have certainly that:
\be
\delta_{\lambda}e = 0 \, \, \, \delta_{e} \lambda = 0 \, \, \, [\delta_{e}, \delta_{\lambda}] = 0
\ee
This allows us to prove the first important result:\\
\\
{\bf Lemma 2} \emph{If the action density is homogeneous in} $\lambda$ \emph{of order} $n$ \emph{Then we have the following relation between the Cartan equation and the Einstein equations}
\be
\lambda \wedge \frac{\delta}{\delta e} Cartan(\lambda, e) = n Einstein (\lambda ,e)
\ee
\bigskip
{\bf Proof}\\
By definition we have:
\be
Cartan(\lambda ,e) = \frac{\delta}{\delta \lambda} L(\lambda,e)
\ee
Since $L(\lambda,e)$ is homogeneous of order $n$ we can write:
\be
\lambda \wedge \frac{\delta}{\delta \lambda} L(\lambda,e) = n L(\lambda,e)
\ee
or
\be
\lambda \wedge Cartan(\lambda,e) = n L(\lambda,e)
\ee
By considering the coframe variation and using {\bf Lemma 1} we get the result.\\
\\
Suppose then that ${\tilde{\lambda}}_{ab}$ is the solution of the Cartan equation; substituting $\lambda = \tilde{\lambda}$ in (11) we get:
\be
\tilde{\lambda} \wedge [\frac{\delta}{\delta e} Cartan(\lambda ,e)]_{\lambda = \tilde{\lambda}} = n Einstein(\tilde{\lambda} ,e)
\ee
In general there is no relation between $\frac{\delta}{\delta e} Cartan(\tilde{\lambda},e)$ and $[\frac{\delta}{\delta e} Cartan(\lambda,e)]_{\lambda = \tilde{\lambda}}$.\\
However if $L(\lambda,e)$ is homogeneous of order $n \ge 2$ in $\lambda,e$ we have:\\
\\
{\bf Lemma 3 :}\\
\\
\emph{If} $L(\lambda,e)$ \emph{is at least quadratic in} $\lambda$ \emph{and} $(\lambda, e)$ \emph{are the only independent variables then:}
\begin{eqnarray}
\tilde{\lambda} \wedge \frac{\delta}{\delta e} Cartan(\tilde{\lambda},e) = \\ \nonumber
\tilde{\lambda} \wedge [\frac{\delta}{\delta e} Cartan(\lambda,e)]_{\lambda = \tilde{\lambda}} 
\end{eqnarray}
\bigskip
{\bf Proof}\\
\\
Consider the expression:
\be
\frac{\delta}{\delta e} Cartan(\tilde{\lambda},e)
\ee
In general $\tilde{\lambda}$ will depend on the coframe so the derivative with respect to the coframe has to be written as:
\begin{eqnarray}
\frac{\delta}{\delta e} Cartan(\tilde{\lambda (e)},e) = \frac{\delta}{\delta e} [Cartan(\lambda,e)]_{\lambda = \tilde{\lambda}} + \\ \nonumber
\frac{\delta}{\delta \tilde{\lambda}} Cartan(\tilde{\lambda},e)\frac{\delta \tilde{\lambda}}{\delta e}
\end{eqnarray}
We can write then:
\begin{eqnarray}
\tilde{\lambda} \wedge \frac{\delta}{\delta e} Cartan(\tilde{\lambda (e)},e) = \tilde{\lambda} \wedge \frac{\delta}{\delta e} [Cartan(\lambda,e)]_{\lambda = \tilde{\lambda}} + \\ \nonumber
\tilde{\lambda} \wedge \frac{\delta}{\delta \tilde{\lambda}} Cartan(\tilde{\lambda},e)\frac{\delta \tilde{\lambda}}{\delta e}
\end{eqnarray} 
If we suppose $L(\lambda,e)$ to be homogeneous of order $ n \ge 2$ we get:
\be
\tilde{\lambda} \wedge \frac{\delta}{\delta \tilde{\lambda}} Cartan(\tilde{\lambda},e) = (n-1) Cartan(\tilde{\lambda},e) 
\ee
Then we conclude that since $Cartan(\tilde{\lambda},e) = 0$, the term on the right hand side of (20) vanishes so that (19) reduces to:
\begin{eqnarray}
\tilde{\lambda} \wedge \frac{\delta}{\delta e} Cartan(\tilde{\lambda},e) = \\ \nonumber
\tilde{\lambda} \wedge [\frac{\delta}{\delta e} Cartan(\lambda,e)]_{\lambda = \tilde{\lambda}} 
\end{eqnarray}
The Lemma is proved.\\
\\
Relation (15) can be rewritten as:
\be
n Einstein(\tilde{\lambda},e) = \tilde{\lambda} \wedge \frac{\delta}{\delta e} Cartan(\tilde{\lambda},e)
\ee
Again:
\be
Cartan(\tilde{\lambda},e) = 0
\ee
so that we get:
\be
Einstein(\tilde{\lambda},e) = 0
\ee
Theorem 1 is then proved.\\
\\
The proven theorem though important somehow restricts the triviality result to the case of homogeneous $L(\lambda,e)$.\\
It is important to observe the fundamental role played by the Cartan equation in the proof of the theorem.\\
This will permits to get a more general result.\\
To obtain that we need the following:\\
\\
{\bf Definition 2 :} \emph{ A Lagrangian density associated with a certain model is said to be QUASI-HOMOGENEOUS if the Cartan equation can be decomposed 
in two or more independent equations of which at least one\\
 is homogeneous of order} $ n \ge 2$ \\
\\
For example when performing the connection variation we may consider two independent equations, one obtained from the diagonal part and another from the traceless part.\\
These two equations are independent and it may happen that a representation can be found in which one of them is homogeneous and the other is not.\\
In this case the model is said to be QUASI-HOMOGENEOUS.\\
Let us call $Cartan^{h}_{1}(\lambda,e) = 0$ the homogeneous one, then we can write:
\be
\lambda \wedge Cartan^{h}_{1}(\lambda,e) = n L_{1}(\lambda,e)
\ee
where in general:
\be
L_{1}(\lambda,e) \neq L(\lambda,e)
\ee
Repeating the steps of the proof of theorem 1 we can prove that $L_{1}(\tilde{\lambda},e) = 0$ and so we get the following:\\
\\
{\bf Theorem 2 :}\\
\\
\emph{Suppose we start from a quasi-homogeneous Lagrangian density} $L(\lambda,e)$ where $Cartan^{h}_{1}(\lambda,e) = 0$ \emph{is the homogeneous part of the Cartan equation, then the generalised Einstein equations are equivalent to:}
\be
L(\tilde{\lambda},e) -  \mu L_{1}(\tilde{\lambda},e) = 0
\ee
\emph{where} $L_{1}(\lambda,e)$ \emph{is defined by:}
\be
\lambda \wedge Cartan^{h}_{1}(\lambda,e) = n L_{1}(\lambda,e)
\ee
\emph{and} $\mu$ \emph{is an arbitrary constant.}\\
\\
\\
{\bf Proof:}\\
\\
The only thing we need is the decomposition: 
\be
L(\lambda,e) =  \mu L_{1}(\lambda,e) + [L(\lambda,e) - \mu L_{1}(\lambda,e)]
\ee
and to apply Theorem 1 to $Cartan^{h}_{1}(\lambda,e)$ and $L_{1}(\lambda,e)$.\\
\\
The proof of theorem 1 and 2 did not depend on the star structure of $L(\lambda,e)$ so we can have any dependence on $\star$; in particular we can consider parity violating terms which do not contain any star (or an even number of them).\\
Consider for example the Lagrangian density in 4 dimensions:
\be
L(\lambda,e) = k R \star 1 + \alpha R^{ab} \wedge e_{a} \wedge e_{b} + \beta T^{a} \wedge T_{a} + \gamma Q_{ab} \wedge e^{a} \wedge T^{b}
\ee
The last three terms do not contain any star so they violate the Parity.\\
By using Theorem 1 we can certainly conclude that the generalised Einstein equations reduce to the Riemannian Einstein equations of general relativity.\\
An interesting case occurs if we consider an action density of the form:
\be
L(\lambda,e) = k R \star 1 + \underbrace{H(\lambda,e)}_{1 \star} + \underbrace{G(\lambda,e)}_{no \, star}
\ee
With $H(\lambda,e)$ and $G(\lambda,e)$ both homogeneous of order two in $\lambda$.\\ 
Again by using Theorem 1 we conclude that:
\be
Einstein(\tilde{\lambda},e) \equiv Einstein(\stackrel{o}{R} \star 1)
\ee
This means that the parity violating terms cancel each other or put in another way the cancellation of $H(\lambda,e)$ and $G(\lambda,e)$ occur independently.\\
In particular consider the action density:
\be
L(\lambda,e) = k R \star 1 + \frac{\alpha}{2}(dQ \wedge \star dQ) + \frac{\beta}{2}(Q \wedge \star Q) + H(\lambda,e) + G(\lambda,e)
\ee
The Cartan equation will separate in two independent parts, the parity invariant part and the parity non-invariant part.\\
The parity invariant part will coincide with the Cartan equation of the action density without the term $G(\lambda,e)$.\\
From the parity invariant part of the Cartan equation I can get the same properties for the non-Riemannian part of the connection.\\ 
Since as seen in Theorem 1,2 the cancellation properties are intimately related to the properties of the Cartan equation, we can certainly state that we get the Proca equation for $Q$ [4,5]:
\be
\alpha d \star d Q + \beta_{0} \star Q = 0
\ee
And the same cancellation of the parity invariant terms will occur.\\
The generalised Einstein equations will reduce to the generalised Einstein equations of the parity invariant theory that is an Einstein-Proca theory in which $Q$ assumes the role of the Proca field.\\
The parity non invariant Cartan equation will be either satisfied or trivial. \footnote{There may be some constraints between the coupling constants appearing in the parity non invariant terms. We are not going to consider this problem in this paper.}
In both cases the parity non invariant terms will give no contribution to the generalised Einstein equations.\\
This gives us the following generalization of the Dereli-Obukhov-Tucker-Wang theorem:\\
\\
{\bf Theorem 3 :} \\
\\
\emph{Consider the action density}
\be
L(\lambda,e) = \kappa  R \star 1 + \frac{\alpha}{2} (dQ \wedge \star dQ) + \frac{\beta}{2}(Q \wedge \star Q) + M(\lambda,e)
\ee
\emph{Where} $M(\lambda,e)$  \emph{is quadratic in} $\lambda$ \emph{but has not definite symmetry in} $\star$ \emph{Then the generalised Einstein equations reduce to the equations of the Einstein-Proca theory.}\\
\section{Concluding Remarks}
\bigskip
We proved that the remarkable properties of cancellation which occur in the generalised Einstein equations of certain non-Riemannian models of gravitation can be connected to the homogeneity properties of the action density with respect to the non-Riemannian part of the connection. In particular if the action density is composed only of homogeneous terms then the Einstein equations become trivial.\\
The most usual case however is the combination of homogeneous and non-homogeneous terms, or the combination of terms with different order of homogeneity. This means that the cancellation can be obtained only for part of the action. The remaining part is the \emph{effective} action density.\\
We also showed that since the simplification properties are related to homogeneous behavior with respect to the non-Riemannian part of the connection; the same result can be obtained if we consider theories which do not preserve parity. This provides a nice extension of the Dereli-Obukhov-Tucker-Wang theorem, to parity non invariant theories.
In this paper the simple case in which the action depends only on the variables $\lambda,e$ has been presented. If other variables are introduced like metric invariant fields, Dilatons, [7-9] etc; the theory becomes more complicate. We expect to study these more general cases in th near future.\\
\\
{\bf Acknowledgments}\\
\\
I wish to thank the International center for cultural cooperation (NOOPOLIS) Italy, for partial financial support.
\newpage
\begin{center}
{\bf REFERENCES}\\
\end{center}
1] F. W. Hehl, J. D. McCrea, E. W. Mielke, Y. Neeman, Phys. Rep. {\bf 258} 1 (1995), and F. Gronwald, Int. J. Mod. Phys. D, {\bf 6} 263 (1997).\\
\\
2] R. W. Tucker, C. Wang, Class. Quant. Grav. {\bf 12} 2587 (1995).\\
\\
3] R. Tucker. C. Wang, Non-Riemannian Gravitational Interactions, Institute of Mathematics, Banach Center Publications, Vol. 41, Warzawa (1997).\\
\\
4] T. Dereli, M. Onder, J. Schray. R. Tucker, C. Wang, Class. Quant. Grav. {\bf 13} L 103 (1996) .\\
\\
5] Y. Obukhov, E. Vlachynsky, W. Esser, F. Hehl, Phys. Rev. D {\bf 56} 12, 776 (1997).\\
\\
6] F. W. Hehl, A. Macias, Int. J. Mod. Phys. D, {\bf 8} 4, (1999) 399.\\
\\
7] R. Scipioni, Class. Quant. Grav. {\bf 16} 2471 (1999).\\
\\
8] R. Scipioni, Phys. Lett. A {\bf 259} 2, 104 (1999).\\
\\
9] R. Scipioni, J. Math. Phys. {\bf 41} 5 (2000).

\end{document}